\documentclass[10pt,letterpaper]{article}
\pdfoutput=1 

\usepackage[utf8]{inputenc} 
\usepackage{lipsum} 

\usepackage[top=2.5cm, bottom=3cm, left=3.5cm, right=3.5cm,
           heightrounded,marginparwidth=1.5cm, marginparsep=1cm]{geometry} 

\usepackage{changepage} 

\usepackage[shortlabels]{enumitem} 

\usepackage[square,numbers,merge,comma,sort&compress]{natbib} 
\makeatletter
\def\NAT@spacechar{\,}  
\makeatother

\usepackage{amsmath,amssymb,amsfonts,amsthm,amsbsy}
\usepackage{mathtools} 

\usepackage{fnpct}
\setfnpct{dont-mess-around} 
\usepackage{slashed,cancel}
\usepackage{old-arrows} 
\usepackage{comment}   
\usepackage{relsize}   
\usepackage{setspace}  
\usepackage{moresize}  
\usepackage{epsfig}
\usepackage{latexsym}
\usepackage{mathrsfs,calligra,aurical} 
\usepackage{calc}
\usepackage{float}
\usepackage{appendix}
\usepackage{bigints}
\usepackage{xargs}
\usepackage{extarrows}
\usepackage{empheq}
\usepackage{soul} 

\usepackage[svgnames]{xcolor}  
\definecolor{Green}{rgb}{0.05, 0.45, 0.25}
\xdefinecolor{dogwoodrose}{rgb}{0.8, 0.1, 0.55}
\xdefinecolor{RRed}{rgb}{0.7, 0.1, 0.525}

\usepackage{tikz}
\usetikzlibrary{scopes,decorations.pathmorphing,patterns,calc,arrows,
                shapes.geometric,shapes.arrows,decorations.markings,plotmarks}
\usepackage{pgfplots}

\usepackage[blocks]{authblk}  

\usepackage[fulladjust]{marginnote}%

\usepackage{graphicx} 
\graphicspath{{Figures/}} 

\usepackage[labelsep=colon]{caption}  
\usepackage[labelsep=colon,aboveskip=10pt,belowskip=10pt]{subcaption}
\DeclareCaptionSubType * [roman]{table}
\usepackage{array,multirow,makecell,booktabs}  
\newcolumntype{M}[2]{>{\centering\arraybackslash$}#1{#2\linewidth}<{$}}
\newcolumntype{T}[2]{>{\centering\arraybackslash}#1{#2\linewidth}<{}}
\newcolumntype{R}[1]{>{\raggedleft\arraybackslash}m{#1\linewidth}<{}}
\newcolumntype{L}[1]{>{\raggedright\arraybackslash}m{#1\linewidth}<{}}
\newcommand\thinrule{\midrule[0.00001pt]}

\makeatletter
\renewcommand\mcell@classz{\@classx
   \@tempcnta \count@
   \prepnext@tok
   \@addtopreamble{
      \ifcase\@chnum
         \hfil
         \mcell@agape{\d@llarbegin\insert@column\d@llarend}\hfil \or
         \hskip1sp
         \mcell@agape{\d@llarbegin\insert@column\d@llarend}\hfil \or
         \hfil\hskip1sp
         \mcell@agape{\d@llarbegin \insert@column\d@llarend}\or
         \mcell@agape{$\vcenter
         \@startpbox{\@nextchar}\insert@column\@endpbox$}\or
         \mcell@agape{\vtop
         \@startpbox{\@nextchar}\insert@column\@endpbox}\or
         \mcell@agape{\vbox
         \@startpbox{\@nextchar}\insert@column\@endpbox}%
      \fi
      \global\let\mcell@left\relax\global\let\mcell@right\relax
    }\prepnext@tok}
\makeatother

\usepackage{titlesec}

\titleformat{\section}{\normalfont\fontsize{11}{11}\bfseries}{\thesection}{0.5em}{}
\titleformat{\subsection}{\normalfont\normalsize\bfseries}{\thesubsection}{0.5em}{}
\titleformat{\subsubsection}{\normalfont\normalsize\bfseries}{\thesubsubsection}{0.5em}{}

\titlespacing*{\section}{0pt}%
                {4ex plus 1ex minus .5ex}{1.75ex plus .25ex minus .25ex}
\titlespacing*{\subsection}{0pt}%
                {3.5ex plus 1ex minus .5ex}{1.25ex plus .2ex minus .2ex}
\titlespacing*{\subsubsection}{0pt}%
                {2.5ex plus 0.75ex minus .2ex}{0.75ex plus .15ex minus .15ex}
\titlespacing*{\paragraph}{0pt}%
                {1.85ex plus 0.5ex minus .15ex}{1em}

\usepackage{titletoc}

\addtocontents{toc}{\addvspace{-0.75em}}  

\titlecontents{section}
  [1.25em] {\addvspace{0.7em plus 0pt}\small}
  {\thecontentslabel\hspace{0.75em}}{}
  {\hspace{0.5em}\titlerule*[0.5em]{.}\contentspage}
  [\addvspace{0.0em plus 0pt}]

\titlecontents{subsection}
  [2.75em] {\addvspace{0.075em plus0pt}\fns}
  {\thecontentslabel\hspace{0.75em}}{\thecontentslabel\hspace{0.75em}}
  {\hspace{0.5em}\titlerule*[0.5em]{.}\small\contentspage}
  [\addvspace{0.075em plus 0pt}]

\usepackage{environ}
\makeatletter
\NewEnviron{subalign}[1]{
\begin{subequations}\label{#1}
\begin{align} \BODY \end{align}
\end{subequations}      }
\makeatother
\makeatletter
\newenvironment{subeqs}%
{\begingroup%
\setlength{\abovedisplayskip}{10pt plus 4pt minus 9pt}%
\setlength{\abovedisplayshortskip}{0pt plus 2pt minus 2pt}%
\setlength{\belowdisplayskip}{12pt plus 3pt minus 9pt}%
\setlength{\belowdisplayshortskip}{7pt plus 3pt minus 4pt}%
\begin{subequations}%
}%
{\end{subequations}\ignorespacesafterend%
\endgroup}%
\makeatother

\makeatletter
{\begingroup%
\setlength{\abovedisplayskip}{{#1}pt plus 3pt minus 3pt}%
\setlength{\abovedisplayshortskip}{{#2}pt plus 3pt}%
\setlength{\belowdisplayskip}{{#3}pt plus 3pt minus 3pt}%
\setlength{\belowdisplayshortskip}{{#4}pt plus 3pt minus 3pt}%
\begin{subequations}%
}%
{\end{subequations}\ignorespacesafterend%
\endgroup}%
\makeatother

\makeatletter
{\begingroup%
\setlength{\abovedisplayskip}{{#1}pt plus 3pt minus 3pt}%
\setlength{\abovedisplayshortskip}{{#2}pt plus 3pt}%
\setlength{\belowdisplayskip}{{#3}pt plus 3pt minus 3pt}%
\setlength{\belowdisplayshortskip}{{#4}pt plus 3pt minus 3pt}%
\begin{equation}%
}%
{\end{equation}\ignorespacesafterend%
\endgroup}%
\makeatother

\makeatletter
{%
\begin{equation}%
\begin{split}%
}%
{\end{split}%
\end{equation}\ignorespacesafterend%
}%
\makeatother

\usepackage{bm}  
\usepackage{dsfont}  

\DeclareMathAlphabet{\mathpzc}{OT1}{pzc}{m}{it}
\DeclareMathAlphabet{\mathcal}{OMS}{cmsy}{m}{n}
\DeclareSymbolFontAlphabet{\Scr}{rsfs}
\DeclareMathAlphabet{\mathbold}{U}{BOONDOX-ds}{m}{n}
\SetMathAlphabet{\mathbold}{bold}{U}{BOONDOX-ds}{b}{n}
\DeclareMathAlphabet{\mathcalboondox}{U}{BOONDOX-calo}{m}{n}
\SetMathAlphabet{\mathcalboondox}{bold}{U}{BOONDOX-calo}{b}{n}
\DeclareMathAlphabet{\mathbcalboondox}{U}{BOONDOX-calo}{b}{n}

\DeclareFontFamily{U}{matha}{\hyphenchar\font45}
\DeclareFontShape{U}{matha}{m}{n}{ <5> <6> <7> <8> <9> <10> gen * matha
                    <10.95> matha10 <12> <14.4> <17.28> <20.74> <24.88> matha12}{}
\DeclareSymbolFont{matha}{U}{matha}{m}{n}
\DeclareMathSymbol{\varleftarrow}{3}{matha}{"D0}
\DeclareMathSymbol{\varrightarrow}{3}{matha}{"D1}
\DeclareMathSymbol{\simeq}{3}{matha}{"14}
\DeclareMathSymbol{\sim}{3}{matha}{"12}
\DeclareMathSymbol{\ll}{3}{matha}{"21}

\newcommand\linkcol{RRed}



\usepackage[breaklinks=true,backref=page]{hyperref}
\hypersetup{
    bookmarks=true,         
    bookmarksnumbered=true,
    pdftoolbar=true,        
    pdfmenubar=true,        
    pdffitwindow=false,     
    pdfauthor={},     
    pdfsubject={},   
    pdfcreator={},   
    pdfproducer={},  
    pdfpagemode={UseNone},
    pdfstartview={FitH},
    colorlinks=true,
    plainpages,
    linktoc=page,
    citecolor=blue,
    filecolor=black,
    linkcolor=\linkcol,
    urlcolor=Green,
}
\renewcommand*{\backref}[1]{}
\renewcommand*{\backrefalt}[4]{%
\ifcase #1 %
\relax
\or
~{\small [\textsc{p.~\fns{\!#2}}]}
\else
~{\small [\textsc{p.~\fns{\!#2}}]}%
\fi}

\usepackage{footnotebackref}
\usepackage{hypernat} 

\usepackage{cleveref} 

\makeatletter
\normalsize
\makeatother

\def\+{~+~}
\def\-{~-~}
\def\={\:=\:}
\newcommand\fns{\footnotesize}

\newcommand\qqquad{\quad\quad\quad}

\newcommand\eps{\varepsilon}

\newcommand\w{\omega}

\newcommand\hgamma{\bar{\gamma}}
\newcommand\Tc{T_\textrm{c}}
\newcommand\ux{\vec{u}_x}
\newcommand\uy{\vec{u}_y}
\newcommand\uz{\vec{u}_z}
\newcommand\E{\mathbf{E}}
\newcommand\B{\mathbf{B}}
\newcommand\Bc{B_\textsc{c}}
\newcommand\Bcone{B_{\textsc{c}{}_{1}}}
\newcommand\Bctwo{B_{\textsc{c}{}_{2}}}
\newcommand\A{\mathbf{A}}

\newcommand\jj{\mathbf{j}}

\newcommand\mstar{m_\star}

\newcommand\mt{\mathrm{m}}
\newcommand\cm{\mathrm{cm}}

\newcommand\s{\mathrm{s}}
\newcommand\Kelv{\mathrm{K}}

\newcommand\cbox{c_{{}_\Box}}

\providecommand{\abs}[1]{\left\lvert#1\right\rvert}

\newcommand{\ms}{\mathsmaller}

\newcommand{\dd}{\partial}
\newcommandx{\tcR}[1]{\textcolor{Crimson}{#1}}
\newcommandx{\tts}[1]{\text{\textsmaller{#1}}}
\newcommandx{\mlt}[1]{\mathlarger{\text{#1}}}
\newcommandx{\dt}[1][1=f,usedefault]{\frac{\partial{#1}}{\partial t}}
\newcommandx{\dtau}[1][1=f,usedefault]{\frac{\partial{#1}}{\partial\tau}}
\newcommandx{\dx}[1][1=f,usedefault]{\frac{\partial{#1}}{\partial x}}
\newcommandx{\ddx}[1][1=f,usedefault]{\frac{\partial^2{#1}}{{\partial x}^2}}
\newcommandx{\dm}[1][1=\mu,usedefault]{\partial_{#1}}
\newcommandx{\dmup}[1][1=\mu,usedefault]{\partial^{#1}}
\newcommandx{\subm}[2][1=p,2=A,usedefault]{{#1}_{\!\mathsmaller{#2}}}
\newcommandx{\subt}[2][1=p,2=A,usedefault]{{#1}_\text{\textsmaller{#2}}}
\newcommandx{\supm}[2][1=p,2=A,usedefault]{{#1}^{\!\mathsmaller{#2}}}
\newcommandx{\supt}[2][1=p,2=A,usedefault]{{#1}^\text{\textsmaller{#2}}}
\newcommandx{\subpt}[3][1=p,2=A,3=B,usedefault]{{#1}^\text{\textsmaller{#3}}_\text{\textsmaller{#2}}}
\newcommandx{\subpm}[3][1=p,2=A,3=B,usedefault]{{#1}^{\mathsmaller{#3}}_{\mathsmaller{#2}}}
\newcommandx{\sh}[1][1=\alpha,usedefault]{\sinh\left(#1\right)}
\newcommandx{\ch}[1][1=\alpha,usedefault]{\cosh\left(#1\right)}
\newcommandx{\sech}[1][1=\alpha,usedefault]{\mathrm{sech}\left(#1\right)}
\newcommandx{\cosech}[1][1=\alpha,usedefault]{\mathrm{cosech}\left(#1\right)} \newcommandx{\LCTd}[4][1=\mu,2=\nu,3=\rho,4=\sigma,usedefault]{\eps_{#1#2#3#4}}
\newcommandx{\LCTu}[4][1=\mu,2=\nu,3=\rho,4=\sigma,usedefault]{\eps^{#1#2#3#4}}

\newcommandx{\gmetr}[2][1=\mu,2=\nu,usedefault]{g_{{#1}{#2}}}
\newcommandx{\invgmetr}[2][1=\mu,2=\nu,usedefault]{g^{{#1}{#2}}}
\newcommandx{\spc}[3][1=\mu,2=a,3=b,usedefault]{{\w_{#1}}^{\!\!{#2}{#3}}}
\newcommandx{\Conn}[3][1=\mu,2=\nu,3=\lambda,usedefault]{{\Gamma_{{#1}{#2}}}^{\!\!#3}}
\newcommandx{\viel}[2][1=\mu,2=a,usedefault]{{e_{#1}}^{\!#2}}
\newcommandx{\inviel}[2][1=a,2=\mu,usedefault]{{e_{#1}}^{#2}}
\newcommandx{\vieluu}[2][1=\mu,2=a,usedefault]{e^{#1#2}}
\newcommandx{\Rdduu}[4][1=\mu,2=\nu,3=a,4=b,usedefault]{{R_{{#1}{#2}}}^{{#3}{#4}}}
\newcommandx{\hgamui}[1][1=0,usedefault]{\hgamma^{\mathsmaller{#1}}}
\newcommandx{\hgamdi}[1][1=0,usedefault]{\hgamma_{{}_{#1}}}
\newcommandx{\gamui}[1][1=0,usedefault]{\gamma^{\mathsmaller{#1}}}
\newcommandx{\gamdi}[1][1=0,usedefault]{\gamma_{{}_{#1}}}

\newcommandx{\emetr}[2][1=\mu,2=\nu,usedefault]{\eta_{{#1}{#2}}}
\newcommandx{\invemetr}[2][1=\mu,2=\nu,usedefault]{\eta^{{#1}{#2}}}
\newcommandx{\hmetr}[2][1=\mu,2=\nu,usedefault]{h_{{#1}{#2}}}
\newcommandx{\invhmetr}[2][1=\mu,2=\nu,usedefault]{h^{{#1}{#2}}}
\newcommandx{\bhmetr}[2][1=\mu,2=\nu,usedefault]{\bar{h}_{{#1}{#2}}}
\newcommandx{\binvhmetr}[2][1=\mu,2=\nu,usedefault]{\bar{h}^{{#1}{#2}}}
\newcommandx{\hud}[2][1=\mu,2=\nu,usedefault]{{h^{#1}}_{\!\!#2}}
\newcommandx{\Ruddd}[4][1=\sigma,2=\mu,3=\lambda,4=\nu,usedefault]{{R^{#1}}_{\!{#2}{#3}{#4}}}
\newcommandx{\Gam}[3][1=\lambda,2=\mu,3=\nu,usedefault]{{\Gamma^{#1}}_{\!{#2}{#3}}}
\newcommandx{\Gamd}[3][1=\mu,2=\nu,3=\lambda,usedefault]{\Gamma_{{#1}{#2}{#3}}}
\newcommandx{\Ricci}[2][1=\mu,2=\nu,usedefault]{R_{{#1}{#2}}}
\newcommandx{\GEinst}[2][1=\mu,2=\nu,usedefault]{G^{{}^\tts{(E)}}_{{#1}{#2}}}
\newcommandx{\Gscr}[3][1=\mu,2=\nu,3=\rho,usedefault]{\mathscr{G}_{{#1}{#2}{#3}}}

\DeclareFixedFont\trfont{OT1}{phv}{b}{sc}{11}

\title{%
       \vspace{-1.0cm}
       \centering\boldmath\LARGE\bfseries%
       Superconductor in static gravitational, electric and magnetic fields with vortex lattice
       \bigskip
       }

\author{\textsc{Giovanni Alberto Ummarino}
\vspace{0.1em}}
\affil{%
{Politecnico di Torino, Dipartimento di Scienza Applicata e Tecnologia, corso Duca degli Abruzzi 24, 10129 Torino, Italy}%
}
\affil{National Research Nuclear University MEPhI, Kashirskoe hwy 31, 115409
       Moscow, Russia%
\vspace{-0.025em}
}%
\affil{\href{mailto:giovanni.ummarino@polito.it}{\texttt{giovanni.ummarino@polito.it}}
       }

\smallskip

\author{\textsc{Antonio Gallerati}
\vspace{0.1em}
}
\affil{%
{Politecnico di Torino, Dipartimento di Scienza Applicata e Tecnologia, corso Duca degli Abruzzi 24, 10129 Torino, Italy}
       }
\affil{Istituto Nazionale di Fisica Nucleare, Sezione di Torino, via Pietro
       Giuria 1, 10125 Torino, Italy%
\vspace{-0.025em}
       }%
\affil{\href{mailto:antonio.gallerati@polito.it}{\texttt{antonio.gallerati@polito.it}}
      }

\date{}

\makeatletter
\patchcmd{\@maketitle}{\begin{center}}{\begin{adjustwidth}{-0.25in}{-0.25in}\begin{center}}{}{}
\patchcmd{\@maketitle}{\end{center}}{\end{center}\end{adjustwidth}}{}{}
\makeatother

\begin{document}

\maketitle
\smallskip

\begin{abstract}
We estimate the conjectured interaction between the Earth’s gravitational field and a superconductor immersed in external, static electric and magnetic field. The latter is close to the sample upper critical field and generates the presence of a vortex lattice. The proposed interaction could lead to multiple, measurable effects. First of all, a local affection of the gravitational field inside the superconductor could take place. Second, a new component of a generalized electric field parallel to the superconductor surface is generated inside the sample.\par
The analysis is performed by using the time-dependent Ginzburg–Landau theory combined with the gravito-Maxwell formalism. This approach leads us to analytic solutions of the problem, also providing the average values of the generated fields and corrections inside the sample. We will also study which are the physical parameters to optimize and, in turn, the most suitable materials to maximize the effect.
\end{abstract}

\bigskip

\tableofcontents
\pagebreak



\section{Introduction} \label{sec:Intro}
The possible interaction between superconductors and gravitational field is an intriguing field of research, providing an interesting connection between condensed matter systems and gravitational interaction, with beneficial effects both in theoretical and applied physics. The seminal paper \cite{DeWitt:1966yi}
set the stage for a deeper analysis of the phenomenon, while, in the following years, a certain amount of scientific literature on the subject was produced
\cite{papini1967detection,Papini:1970cw,hirakawa1975superconductors,
anandan1984relthe,ross1983london,dinariev1987relativistic,
peng1990new,peng1991electrodynamics,peng1991interaction,
li1991effects}. 
The underlying idea behind this line of research is that, under certain conditions, a gravity/supercondensate interplay should exist, resulting in a slight affection of the local gravitational field through the interaction with suitable condensate systems.
Finally, in 1992 Podkletnov and Nieminen proposed a laboratory experimental configuration to detect the conjectured  mutual interplay \cite{podkletnov1992possibility,podkletnov1997weak}. Due the complexity of the experimental setting and the high costs involved, the experiment was then repeated only in simplified versions \cite{hathaway2003gravity}, without achieving conclusive results.\par
After the above experiments, a series of theoretical explanations have been proposed in the literature  \cite{li1997static,Modanese:2001wv,Modanese:1996sq,Wu:2003aq}, including the first and most convincing interpretation \cite{Modanese:1995tx,Modanese:1996zm} given in terms of a coupling of the gravitational field with an unconventional state of matter (superfluid) in a quantum gravity framework. In all these works, the contribution of the superfluid is included in the energy momentum tensor, exploiting the formalism of general relativity. Unfortunately, this approach does not allow to extrapolate quantitative predictions to be put in relation with possible laboratory experiments: one is then led to consider alternative, phenomenological evidences to better understand the proposed interplay.
\par\smallskip
Parallel to DeWitt (and subsequent) studies about the coupling between supercondensates and gravity, other theoretical \cite{schiff1966gravitation,Dessler:1968zz,torr1993gravitoelectric,mashhoon2000gravitoelectromagnetism,mashhoon2007review} and experimental \cite{witteborn1967experimental,witteborn1968experiments,herring1968gravitationally} researches were conducted about generalized electric-type fields induced in (super)conductors by the presence of the Earth's weak gravitational field. The main result of those studies was the introduction of a generalized electric-type field, characterized by an electrical component and a gravitational one, leading to detectable corrections to the free fall of charged particles. These results then led to the definition of a set of generalized, fundamental fields featuring both gravitational and electromagnetic components \cite{agop2000local,agop2000some,Ruggiero:2002hz,Tartaglia:2003wx,Ummarino:2017bvz,Behera:2017voq}. This gravito-Maxwell formalism, valid in the weak gravity regime, is particularly suitable for treating the behaviour of a superfluid immersed in the Earth's gravitational field \cite{Ummarino:2019cvw,Ummarino:2020loo,Ummarino:2021vwc,Gallerati:2021ops}.\par
In this paper we again use the time-dependent Ginzburg Landau theory combined with the gravito-Maxwell formulation. With respect to our previous analyses, this time we also consider the presence of external static electric and magnetic fields. In particular, the magnetic field value is very close to the critical magnetic field of the superconductor, and its presence also determines the presence of a vortex lattice. These new ingredients can lead to an enhancement of the interaction with the gravitational field, once chosen appropriate sample parameters and geometry.


\section{Gravito-Maxwell formalism}%
\label{sec:gravMax}
Here we consider a weak gravitational background, where the (nearly-flat) spacetime metric $g_{\mu\nu}$ is expressed as
\begin{equation}
g_{\mu\nu}\:\simeq\:\eta_{\mu\nu}+h_{\mu\nu}\:,
\end{equation}
with $\eta_{\mu\nu}=\mathrm{diag}(-1,+1,+1,+1)$ and where $h_{\mu\nu}$ is a small perturbation of the flat Minkowski spacetime%
\footnote{%
here we work in the `mostly plus' convention and natural units $c=\hbar=1$
}.
If we now introduce the symmetric traceless tensor
\begin{equation}
\bar{h}_{\mu\nu}\=h_{\mu\nu}-\frac12\,\eta_{\mu\nu}\,h\:,
\end{equation}
with $h=h^\sigma{}_\sigma$, it can be shown that the Einstein equations in the harmonic De Donder gauge $\partial^{\mu}\bar{h}_{\mu\nu}\simeq0$ can be rewritten, in linear approximation, as \cite{Ummarino:2017bvz,Ummarino:2019cvw,Ummarino:2020loo,Ummarino:2021vwc,Gallerati:2021ops}
\begin{equation}
R_{\mu\nu}-\frac12\,g_{\mu\nu}\,R\=\partial^{\rho}\mathscr{G}_{\mu\nu\rho}\:=\:8\pi\mathrm{G}\,T_{\mu\nu}\:,
\end{equation}
having also defined the tensor
\begin{equation}
\mathscr{G}_{\mu\nu\rho}\:\equiv\:\partial_{{[}\nu}\bar{h}_{\rho{]}\mu}+\partial^{\sigma}\eta_{\mu{[}\rho}\,\bar{h}_{\nu{]}\sigma}
     \:\simeq\:\partial_{{[}\nu}\bar{h}_{\rho{]}\mu}\:.
\end{equation}

\subsection{Gravito-Maxwell formulation}
We then introduce the fields \cite{agop2000local,Ummarino:2017bvz}
\begin{align}
\mathbf{E}_\textrm{g}=-\frac12\,\mathscr{G}_{00i}=-\frac12\,\partial_{{[}0}\bar{h}_{i{]}0}\,,
\qquad
\mathbf{A}_\textrm{g}=\frac14\,\bar{h}_{0i}\,,
\qquad
\mathbf{B}_\textrm{g}=\frac14\,{\varepsilon_i}^{jk}\,\mathscr{G}_{0jk}\,,
\end{align}
for which we get, restoring physical units, the set of equations \cite{agop2000local,Ummarino:2017bvz,Ummarino:2019cvw,Ruggiero:2002hz,Tartaglia:2003wx,Vieira:2016csi,Behera:2017voq,Sbitnev:2019iyz,Ummarino:2020loo,Giardino:2018ffd,Gallerati:2020tyq,Ummarino:2021vwc,Gallerati:2021ops}
\begin{equation}
\begin{alignedat}{2}
&\nabla\cdot\mathbf{E}_\text{g}=4\pi\mathrm{G}\,\rho_\text{g}\,,\qquad\; &
&\nabla\cdot\mathbf{B}_\text{g}=0 \,,
\\[2.5\jot]
&\nabla\times\mathbf{E}_\text{g}=-\dfrac{\partial\mathbf{B}_\text{g}}{\partial t}\,,\qquad\; &
&\nabla\times\mathbf{B}_\text{g}=4\pi\mathrm{G}\,\frac{1}{c^2}\,\mathbf{j}_\text{g}
    +\frac{1}{c^2}\,\frac{\partial\mathbf{E}_\text{g}}{\partial t}\,,
\end{alignedat}
\end{equation}
having defined the mass density \:$\rho_\text{g}\equiv-T_{00}$\: and the mass current density \:$\mathbf{j}_\text{g}\equiv T_{0i}$\,.
The above equations have the same formal structure of the Maxwell equations, with $\mathbf{E}_\textrm{g}$ and $\mathbf{B}_\textrm{g}$ gravitoelectric and gravitomagnetic field, respectively.

\subsection{Generalized fields and equations}
Now we introduce generalized electric/magnetic fields, scalar and vector potentials, featuring both electromagnetic and gravitational contributions \cite{agop2000local,agop2000some,Ummarino:2017bvz,Ummarino:2019cvw}:
\begin{equation}
\mathbf{E}=\mathbf{E}_\text{e}+\frac{m}{e}\,\mathbf{E}_\text{g}\,,\qquad
\mathbf{B}=\mathbf{B}_\text{e}+\frac{m}{e}\,\mathbf{B}_\text{g}\,,\qquad
\phi=\phi_\text{e}+\frac{m}{e}\,\phi_\text{g}\,,\qquad
\mathbf{A}=\mathbf{A}_\text{e}+\frac{m}{e}\,\mathbf{A}_\text{g}\,,
\label{eq:genfields}
\end{equation}
where $m$ and $e$ identify the mass and electronic charge, respectively.
The generalized Maxwell equations for the new fields read \cite{agop2000local,Ummarino:2017bvz,Ummarino:2019cvw,Behera:2017voq,Ummarino:2020loo,Ummarino:2021vwc,Gallerati:2021ops}:
\begin{equation}
\begin{alignedat}{2}
&\nabla\cdot\mathbf{E}=\left(\frac{1}{\varepsilon_0}+\frac{1}{\varepsilon_\text{g}}\right)\rho\,,\qquad\;&
&\nabla\cdot\mathbf{B}=0 \,,
\\[2.5\jot]
&\nabla\times\mathbf{E}=-\dfrac{\partial\mathbf{B}}{\partial t}\,,\qquad\;&
&\nabla\times\mathbf{B}=(\mu_0+\mu_\text{g})\,\mathbf{j}
    +\frac{1}{c^2}\,\frac{\partial\mathbf{E}}{\partial t}\,,
\end{alignedat}
\end{equation}
where $\varepsilon_0$ and $\mu_0$ are the vacuum electric permittivity and magnetic permeability. In the above equations, $\rho$ and $\mathbf{j}$ identify the electric charge density and electric current density, respectively, while the mass density and the mass current density vector can be expressed in terms of the latter as
\begin{equation}
\rho_\text{g}=\frac{m}{e}\,\rho\,,\qquad\quad
\mathbf{j}_\text{g}=\frac{m}{e}\:\mathbf{j}\:,
\end{equation}
while the vacuum \emph{gravitational} permittivity $\varepsilon_\text{g}$ and permeability $\mu_\text{g}$ read
\begin{equation}
\varepsilon_\text{g}=\frac{1}{4\pi\mathrm{G}}\,\frac{e^2}{m^2}\:,\qquad\quad
\mu_\text{g}=\frac{4\pi\mathrm{G}}{c^2}\,\frac{m^2}{e^2}\:.
\end{equation}


\section{The model}

\subsection{Time-dependent Ginzburg--Landau formulation}
\begin{figure}[!htp]
\captionsetup{skip=5pt,belowskip=15pt,font=small,labelfont=small,format=hang}
\centering
\includegraphics[width=0.85\textwidth,keepaspectratio]{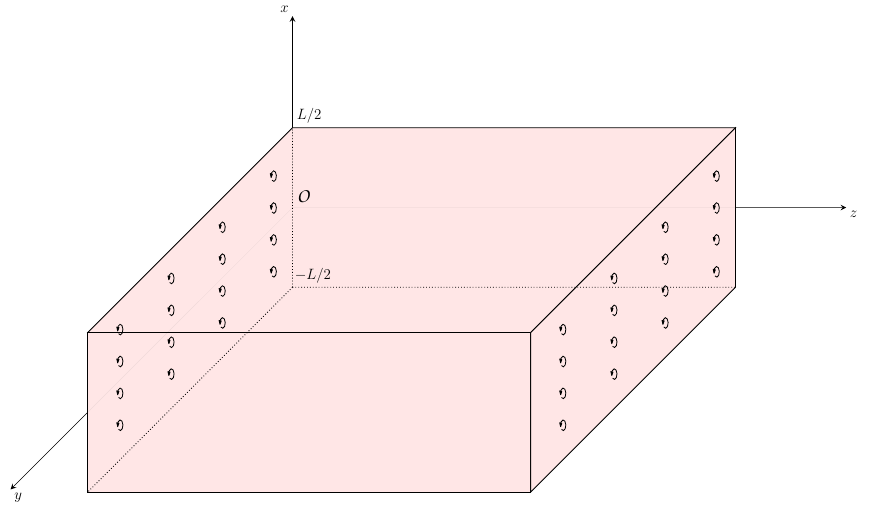}
\caption{Physical setup.}
\label{fig:sample}
\end{figure}

The time-dependent Ginzburg--Landau equations (TDGL) can be written as \cite{tang1995time,lin1997ginzburg,ullah1991effect,ghinovker1999explosive,kopnin1999time,fleckinger1998dynamics,du1996high,oripov2020time}:
\begin{subeqs} \label{eq:TDGL0}
\begin{align}
&\frac{\hbar^2}{2\,\mstar}\left(i\,\nabla
        +\frac{2\,e}{\hbar}\,\A\right)^{\!2}\psi \-a\,\psi\+b\,\abs{\psi}^2\psi
        \;=\,-\frac{\hbar^2}{2\,\mstar\,\mathcal{D}}\left(\frac{\dd}{\dd t}
                +\frac{2\,i\,e}{\hbar}\,\phi\right)\,\psi\;,
\\[2\jot]
&\nabla\times\nabla\times\A\-\nabla\times\B_0\=\mu_0\,\big(\jj_\text{n}+\jj_\text{s}\big)\,,
\end{align}
\end{subeqs}
where $\B_0$ is the external magnetic field, $\mathcal{D}$ is the diffusion coefficient, $\sigma$ the conductivity in the normal phase and
\begin{equation}
a\=a(T)\=a_{0}\,(T-\Tc)\,,\qquad\qquad
b\=b(\Tc)\,,\qquad
\end{equation}
$a_0$, $b$ being positive constants. The contributions related to the normal current and supercurrent densities can be explicitly written as
\begin{equation}
\begin{split}
\jj_\text{n}&\=-\sigma\left(\frac{\dd\A}{\dd t}+\nabla\phi\right)\:,
\\[3\jot]
\jj_\text{s}&\=-i\,\hbar\,\frac{e}{\mstar}\left(\psi^*\,\nabla\psi-\psi\,\nabla\psi^*\right)
    -\frac{4\,e^2}{\mstar}\,\abs{\psi}^2\A\:.
\end{split}
\end{equation}
%
%
Let us put ourselves in the London gauge $\nabla\cdot\A=0$ so that
\begin{equation}
\nabla^2\A\=-\mu_0\left(\jj_\text{n}+\jj_\text{s}\right)\:.
\end{equation}
We now consider a sample of thickness $L$ along the (vertical) $\ux$ direction and very large dimensions along $\uy$, $\uz$, see Fig.\ \ref{fig:sample}. The sample features a square lattice of vortices, whose axes are directed along the direction of the static magnetic field $\B_0$ that we choose to be
\begin{equation}
\B_0=B_0\,\uz\:,
\end{equation}
together with a vector potential $\A$ of the form
\begin{equation}
\A=B_0\,x\,\uy\:.
\end{equation}
We then consider the presence of a constant external (standard) electric field $\E_0^{{}^\mlt{(e)}}$ along the $\ux$ direction, so that a \emph{generalized} static field $\E_0$ can be expressed as
\begin{equation}
\E_0\=\E_0^{{}^\mlt{(e)}}+\E_0^{{}^\mlt{(g)}}\=\left(E_0^{{}^\mlt{(e)}}-E_0^{{}^\mlt{(g)}}\right)\,\ux
    \=\left(E_0^{{}^\mlt{(e)}}-\frac{m}{e}\,g\right)\,\ux\=E_0\,\ux\:,\quad
\end{equation}
with a related scalar potential $\phi_0\=-E_0\,x$.\par
The boundary and initial conditions are
%
\begin{align}
  \left.
  \begin{aligned}
  \left(i\,\nabla\psi+\frac{2\,e}{\hbar}\,\A\,\psi\right)\cdot\mathbf{n}=0&  \cr
  \hfill\nabla\times\A\cdot\mathbf{n}=\B_0\cdot\mathbf{n}&  \\[1.5\jot]
  \hfill\A\cdot\mathbf{n}=0&
  \end{aligned}
  \;\;\right\} \; \text{on }\dd\Omega\times(0,t)\;,
\qquad\quad
  \left.
  \begin{aligned}
  \psi(x,0)&\=\psi_0(x) \cr
  \A(x,0)&\=\A_0(x)
  \end{aligned}
  \!\!\!\!\right\} \; \text{on }\Omega\;,\qquad
\label{eq:boundary}
\end{align}
%
where $\dd\Omega$ is the boundary of a smooth and simply connected domain in $\mathbb{R}^\ms{\textrm{N}}$. We denote by $\A_0$ the external vector potential, coinciding with the internal value for $t\leq0$, i.e.\ when the sample is in the normal state and the material is very weakly diamagnetic. For $t<0$ we also have  $T<\Tc$ and $B>\Bctwo$, while at $t=0$ we still have $T<\Tc$ but $B\simeq\Bctwo$.\par
If we put ourselves very close to $\Bctwo$, it is possible to find a solution of the linearized TDGL equations \cite{kopnin1993flux,kopnin2001theory} for the order parameter of the form
\begin{equation}
\psi(x,y,t)\=
    \sum_{n=-\infty}^{\infty}\! c_n\,\exp\left(i\,q\,n\left(y+\frac{E_0}{B_0}\:t\right)\!\right)\;\:
    \exp\left(-\frac{1}{2\,\xi(T)}\left(x-\frac{\hbar\,q\,n}{2\,e\,B_0}\right)^{\!2}
        +i\,\frac{e\,E_0\,\xi^2(T)}{\hbar\,\mathcal{D}}\left(x-\frac{\hbar\,q\,n}{2\,e\,B_0}\right)\right)\:.
\end{equation}
Since we are very close to $\Bctwo$, this solution of the linearized TDGL equations describes the behaviour of an ordered vortex lattice, moving under the influence of the external electric field%
\footnote{%
we should also note that, although this is an exact solution just below $\Bctwo$, this formula does not necessarily hold for different values of the magnetic field, such as the lower critical field $\Bcone$; indeed, close to $\Bctwo$ the vortices are very close to each other, at distances of the order the coherence length $\xi(T)$ (they are so densely packed that their cores are essentially touching \cite{ketterson1999superconductivity});
}%
\footnote{%
the motion of the vortices under the influence of the external electric field causes dissipative phenomena even in the superconducting state; one way to avoid the effect is to add defects to anchor the vortices (vortex pinning), in order to reduce or eliminate energy dissipation \cite{kopnin1993flux}
}. For a square lattice, $q$ expresses the distance between adjacent vortices \cite{hoffmann2012ginzburg}
\begin{equation}
q\:\simeq\:\frac{2\pi}{\xi(T)}\:,
\end{equation}
and we can also replace the general $c_n$ coefficient with the correspondent $\cbox$ expression for the square lattice:
\begin{equation}
c_n\;\;\rightarrow\;\;\:\cbox=\frac{2\sqrt{2\pi}}{\xi^2(T)}\:,
\end{equation}
the $\cbox$ coefficients being then independent of $n$.

\paragraph{Dimensionless TDGL.}
In order to write eqs.\ \eqref{eq:TDGL0} in a dimensionless form, the following expressions are introduced:
\begingroup%
\setlength{\belowdisplayskip}{2pt plus 3pt minus 4pt}%
\setlength{\belowdisplayshortskip}{2pt plus 3pt minus 4pt}%
\begin{equation} \label{eq:param}
\begin{split}
&\psi_0^2(T)=\frac{\abs{a(T)}}{b}\,,\qquad\;
\xi(T)=\frac{\hbar}{\sqrt{2\,\mstar\,|a(T)|}}\,,\qquad\;
\lambda(T)=\sqrt{\frac{b\,\mstar}{4\mu_0|a(T)|\,e^2}}\,,\qquad\;
\kappa=\frac{\lambda(T)}{\xi(T)}\,,\quad
\\[2\jot]
&\tau(T)=\frac{\lambda^2(T)}{\mathcal{D}}\,,\qquad\;
\eta=\mu_0\,\sigma\,\mathcal{D}\,,\qquad\;\;
\Bc(T)=\sqrt{\frac{\mu_0\,\abs{a(T)}^2}{b}}=
    \frac{\hbar}{2\sqrt{2}\,e\,\lambda(T)\,\xi(T)}\,,
\end{split}
\end{equation}
\endgroup
where $\lambda(T)$, $\xi(T)$ and $\Bc(T)$ are the penetration depth, coherence length and thermodynamic critical field, respectively. We also define the dimensionless quantities
\begin{equation}
t'=\:\frac{t}{\tau}\,,\qqquad
x'=\:\frac{x}{\lambda}\,,\qqquad
y'=\:\frac{y}{\lambda}\,,\qqquad
x_0'=\:\frac{x_0}{\lambda}\,,\qqquad
\psi'=\:\frac{\psi}{\psi_0}\,,
\label{eq:dimlesscoord}
\end{equation}
and the dimensionless potentials, fields and currents can be expressed as:
\begingroup%
\setlength{\belowdisplayskip}{25pt plus 3pt minus 4pt}%
\setlength{\belowdisplayshortskip}{25pt plus 3pt minus 4pt}%
\begin{equation}
\A'=\frac{\A\,\kappa}{\sqrt{2}\,\Bc\,\lambda}\,,\qquad
\phi'=\frac{\phi\,\kappa}{\sqrt{2}\,\Bc\,\mathcal{D}}\,,\qquad
\E'=\frac{\E\,\lambda\,\kappa}{\sqrt{2}\,\Bc\,\mathcal{D}}\,,\qquad
\B'=\frac{\B\,\kappa}{\sqrt{2}\,\Bc}\,,\qquad
\jj'=\frac{\jj\,\mu_0\,\lambda\,\kappa}{\sqrt{2}\,\Bc}\,.
\label{eq:dimlessfields}
\end{equation}
\endgroup
We now consider the dimensionless version of the London gauge, \,$\nabla'\cdot\A'=0$.\, Inserting \cref{eq:dimlesscoord,eq:dimlessfields} in eqs.\ \eqref{eq:TDGL0}, provides the dimensionless TDGL equations (we drop the primes for the sake of notational simplicity) in a bounded, smooth and simply connected domain in $\mathbb{R}^\ms{\textrm{N}}$ \cite{tang1995time,lin1997ginzburg}:
\begin{subeqs}\label{eq:dimlessTDGL}
\begingroup%
\setlength{\abovedisplayskip}{2pt plus 3pt minus 4pt}%
\setlength{\abovedisplayshortskip}{2pt plus 3pt minus 4pt}%
\begin{align}
&\frac{\dd\psi}{\dd t} \,+\, i\,\phi\,\psi
    \,+\,\kappa^2\left(\abs{\psi}^2-1\right)\,\psi
    \,+\,\left(i\,\nabla+\A\right)^2 \psi\=0 \;,
\label{subeq:dimlessTDGLn1}
\\[2\jot]
&\nabla\times\nabla\times\A\,-\,\nabla\times\B_0
    \=\jj_\text{n}+\jj_\text{s}
    \:=\,-\,\eta\,\left(\frac{\dd\A}{\dd t}+\nabla\phi\right)
        -\frac{i}{2}\left(\psi^*\nabla\psi-\psi\nabla\psi^*\right)
        -\abs{\psi}^2\A\,,
\label{subeq:dimlessTDGLn2}
\end{align}
\endgroup
\end{subeqs}
and the boundary and initial conditions \eqref{eq:boundary} become, in the dimensionless form
\begin{align}
  \left.
  \begin{aligned}
  \left(i\,\nabla\psi+\A\,\psi\right)\cdot\mathbf{n}=0&\cr
  \nabla\times\A\cdot\mathbf{n}=\B_0\cdot\mathbf{n}&\cr
  \A\cdot\mathbf{n}=0&
  \end{aligned}
  \!\!\!\right\} \; \text{on }\dd\Omega\times(0,t)\;;
\qquad\quad
  \left.
  \begin{aligned}
  \psi(x,0)&\=\psi_0(x)\cr
  \A(x,0)&\=\A_0(x)
  \end{aligned}
  \!\!\!\!\right\} \; \text{on }\Omega\;.\qquad
\label{eq:dimlessboundary}
\end{align}

\subsection{Solving dimensionless TDGL}
We now discuss how to get an analytic approximate solution to the above dimensionless TDGL.\par
First of all, we write the following first-order expression for the dimensionless order parameter:
\begin{equation}
\psi(x,y,t)\=
    \sum_{n=-\infty}^{\infty}\! \abs{c_n}\,\exp\left(i\,q\,n\left(y+\frac{E_0}{B_0}\:t\right)\!\right)\;\:
    \exp\left(-\frac{\kappa^2}{2}\left(x-n\,x_0\right)^2
        +i\,\frac{E_0}{\kappa}\left(x-n\,x_0\right)\right)\:,
\end{equation}
with
\begin{equation}
\abs{\psi}^2\=\sum_{n=-\infty}^{\infty}\!\abs{c_n}^2\,\exp\left(-\kappa^2\left(x-n\,x_0\right)^2\right)\:.
\end{equation}
The equations for the vector potential components can be written as
\begin{equation} \label{eq:vectpot0}
\begin{split}
\frac{\dd^2 A_x(x,t)}{\dd x^2}&\=
    \eta\,\left(\frac{\dd A_x(x,t)}{\dd t}-E_0\right) +\left(A_x(x,t)-\frac{E_0}{\kappa}\right)
        \sum_{n=-\infty}^{\infty}\!\abs{c_n}^2\,\exp\Big(-\kappa^2(x-n\,x_0)^2\Big)\:,
\\[1ex]
\frac{\dd^2 A_y(x,t)}{\dd x^2}&\=
    \eta\,\frac{\dd A_y(x,t)}{\dd t} +\sum_{n=-\infty}^{\infty}\!\big(A_y(x,t)-2\,\pi\,\kappa\,n\,\big)\,\abs{c_n}^2\,\exp\Big(-\kappa^2(x-n\,x_0)^2\Big)\:,
\\[1ex]
\frac{\dd^2 A_z(x,t)}{\dd x^2}&\=
    \eta\,\frac{\dd A_z(x,t)}{\dd t} +\sum_{n=-\infty}^{\infty}\!\abs{c_n}^2\,\exp\Big(-\kappa^2(x-n\,x_0)^2\Big)\:.
\end{split}
\end{equation}
We now want to obtain explicit solutions for the dimensionless order parameter at first order in $E_0$. To this end, we have to estimate the summations
\begin{equation}
\sum_{n=-\infty}^{\infty}\!\abs{c_n}^2\,\exp\left(-\kappa^2\left(x-n\,x_0\right)^2\right)\,,\qqquad
\sum_{n=-\infty}^{\infty}\!n\,\abs{c_n}^2\,\exp\left(-\kappa^2\left(x-n\,x_0\right)^2\right)\,.
\label{eq:summations}
\end{equation}
Since we are considering a square vortex lattice, we can replace the general coefficients $c_n$ with $\cbox$ that, for the case under consideration, can be expressed as \cite{hoffmann2012ginzburg}
\begin{equation}
\cbox^2\=2\,\sqrt{2\pi}\,\kappa^2\:,
\end{equation}
being then a function of $\kappa$ only. Moreover, we are interested in high-$\Tc$ superconductors, so that we also have large values for the $\kappa$ parameter, $\kappa^2\,\ms{\gtrsim}\,10^4$. This gives us the following estimates for the above \eqref{eq:summations}:
\begin{equation}
\begin{split}
&\sum_{n=-\infty}^{\infty}\!\abs{c_n}^2\,\exp\left(-\kappa^2\left(x-n\,x_0\right)^2\right)
    \=\cbox^2\,e^{-\kappa^2x^2}\sum_{n=-\infty}^{\infty}\!e^{-\kappa^2n^2x_0^2}\;e^{2\,x\,x_0\,n\,\kappa^2}
    \:\simeq\:\cbox^2\,e^{-\kappa^2x^2}\:,
\\[1ex]
&\sum_{n=-\infty}^{\infty}\!n\,\abs{c_n}^2\,\exp\left(-\kappa^2\left(x-n\,x_0\right)^2\right)
    \:\simeq\:0\:,
\end{split}
\end{equation}
where only the $n=0$ term gives a non negligible contribution in the first summation. Equations \eqref{eq:vectpot0} can be rewritten
\begin{equation} \label{eq:vectpot1}
\begin{split}
\frac{\dd^2 A_x(x,t)}{\dd x^2}&\=
    \eta\,\left(\frac{\dd A_x(x,t)}{\dd t}-E_0\right) +\left(A_x(x,t)-\frac{E_0}{\kappa}\right)\,\cbox^2\,e^{-\kappa^2x^2}\:,
\\[1ex]
\frac{\dd^2 A_y(x,t)}{\dd x^2}&\=
    \eta\,\frac{\dd A_y(x,t)}{\dd t}+A_y(x,t)\,\cbox^2\,e^{-\kappa^2x^2}\:,
\\[1ex]
\frac{\dd^2 A_z(x,t)}{\dd x^2}&\=
    \eta\,\frac{\dd A_z(x,t)}{\dd t}+\cbox^2\,e^{-\kappa^2x^2}\:.
\end{split}
\end{equation}
Since we are dealing with materials featuring a very large value for the $\kappa$ parameter, $\kappa^2\,\ms{\gtrsim}\,10^4$, we can also use the approximation
\begin{equation}
e^{-\kappa^2x^2}\:\simeq\:\frac{\sqrt{\pi}}{\kappa}\,\delta(x)\:,
\end{equation}
and our equation for the vector potential components become
\begin{subeqs}\label{eq:vectpot2}
\begin{align}
\frac{\dd A_x(x,t)}{\dd t}\:&\simeq\:
    \frac{1}{\eta}\,\frac{\dd^2 A_x(x,t)}{\dd x^2}
    -\left(A_x(x,t)-\frac{E_0}{\kappa}\right)\,\cbox^2\,\frac{\sqrt{\pi}}{\eta\,\kappa}\,\delta(x)+E_0\:,
\label{subeq:Ax}
\\[1ex]
\frac{\dd A_y(x,t)}{\dd t}\:&\simeq\:
    \frac{1}{\eta}\,\frac{\dd^2 A_y(x,t)}{\dd x^2}
    -A_y(x,t)\:\cbox^2\,\frac{\sqrt{\pi}}{\eta\,\kappa}\,\delta(x)\:,
\label{subeq:Ay}
\\[1ex]
\frac{\dd A_z(x,t)}{\dd t}\:&\simeq\:
    \frac{1}{\eta}\,\frac{\dd^2 A_z(x,t)}{\dd x^2}
    -\cbox^2\,\frac{\sqrt{\pi}}{\eta\,\kappa}\,\delta(x)\:.
\label{subeq:Az}
\end{align}
\end{subeqs}
The initial conditions for the vector potential components are:
%
%
\begin{equation}
A_x\left(x,\,0\right)=0\,,\qquad\;
A_y\left(x,\,0\right)=B_0\,x\,,\qquad\;
A_z\left(x,\,0\right)=0\,,
\label{eq:initcond}
\end{equation}
and the generalized electric field $\E$ inside the superconductor can be obtained from
\begin{equation}
\E\=-\frac{\dd\A}{\dd t}-\nabla\phi\:.
\end{equation}

\paragraph{Averaging over space.}
%
Let us now study the average effects of the generalized electric field inside the superconductor. To this end, we integrate the vector potential components \eqref{eq:vectpot2} over the $x$-variable \cite{sanders2007averaging}.\par
First, we integrate over $x$ eq.\ \eqref{subeq:Az}, for an interval $x\in[-L/2,\,L/2]$ %
\footnote{%
we emphasize that we have dropped the primes for the sake of notational simplicity, so here both $L$ and $x$ correspond to the dimensionless $L'$ and $x'$ of \eqref{eq:dimlesscoord}; in particular, here one would explicitly have for the dimensionless thickness $L'=L/\lambda$,\, $L$ being the physical thickness and $\lambda$ the penetration depth.
}. This gives
\begin{equation}
\frac{\dd\bar{A}_z(t)}{\dd t}
    \=-\cbox^2\,\frac{\sqrt{\pi}}{\eta\,\kappa\,L}\:,
\end{equation}
having defined the averaged component \,$\bar{A}_z(t)=\frac{1}{L}\!\bigintssss\limits_{{}^{-L/2}}^{{}_{L/2}}\!\!dx\,A_z(x,t)$\, and having used symmetric conditions for the first derivatives with respect to $x$. The above equation is easily solved as
\begin{equation}
\bar{A}_z(t)\=-\cbox^2\,\frac{\sqrt{\pi}}{\eta\,\kappa\,L}\,t+\bar{A}_z(0)
    \=-\cbox^2\,\frac{\sqrt{\pi}}{\eta\,\kappa\,L}\,t\:,
\label{eq:barAz}
\end{equation}
having used initial conditions \eqref{eq:initcond} to set $\bar{A}_z(0)=0$. This in turn gives for the averaged, generalized electric field $\bar{E}_z$ component
\begin{equation}
\bar{E}_z\=\cbox^2\,\frac{\sqrt{\pi}}{\eta\,\kappa\, L}
    \=\frac{2\sqrt{2}\,\pi\,\kappa}{\eta\,L}\:.
\end{equation}
The averaged differential equation for the $\bar{A}_y(t)$ component is obtained from \eqref{subeq:Ay} and reads
\begin{equation}
\frac{\dd\bar{A}_y(t)}{\dd t}
    \=-\bar{A}_y(t)\:\cbox^2\,\frac{\sqrt{\pi}}{\eta\,\kappa\,L}\:,
\end{equation}
having used the approximation $A_y(0,t)\simeq \bar{A}_y(t)$. The above equation would give for the averaged component
\begin{equation}
\bar{A}_y(t)\=\bar{A}_y(0)\:\exp\left(-\cbox^2\,\frac{\sqrt{\pi}}{\eta\,\kappa\,L}\:t\right)\=0\:,
\end{equation}
using again initial condition \eqref{eq:initcond}, the resulting field $\bar{E}_y(t)$ being then zero.\par
Finally, for the vertical component we have from \eqref{subeq:Ax}
\begin{equation}
\frac{\dd \bar{A}_x(t)}{\dd t}
    \=-\left(\frac{\bar{A}_x(t)}{L}-\frac{E_0}{\kappa}\right)\,\cbox^2\,\frac{\sqrt{\pi}}{\eta\,\kappa}+E_0\:,
\end{equation}
that is solved, always using the approximation $A_x(0,t)\simeq \bar{A}_x(t)$, for
\begin{equation}
\begin{split}
\bar{A}_x(t)&\=\bar{A}_x(0)\,\exp\left(-\cbox^2\,\frac{\sqrt{\pi}}{\eta\,\kappa\,L}\,t\right)
    +E_0\left(\frac{L}{\kappa}+\frac{\eta\,\kappa\,L}{\cbox^2\sqrt{\pi}}\right)\,
    \left(1-\exp\left(-\cbox^2\,\frac{\sqrt{\pi}}{\eta\,\kappa\,L}\,t\right)\,\right)\=
\\[1ex]
    &\=E_0\left(\frac{L}{\kappa}+\frac{\eta\,\kappa\,L}{\cbox^2\sqrt{\pi}}\right)\,
    \left(1-\exp\left(-\cbox^2\,\frac{\sqrt{\pi}}{\eta\,\kappa\,L}\,t\right)\,\right)
\end{split}
\end{equation}
having again used conditions \eqref{eq:initcond}. We finally get for the averaged $E_x(t)$ component of the generalized electric field
\begin{equation}\label{eq:Excorr}
\begin{split}
\bar{E}_x(t)&\=E_0
    -E_0\left(\frac{L}{\kappa}+\frac{\eta\,\kappa\,L}{\cbox^2\sqrt{\pi}}\right)\,
    \cbox^2\,\frac{\sqrt{\pi}}{\eta\,\kappa\,L}\,\exp\left(-\cbox^2\,\frac{\sqrt{\pi}}{\eta\,\kappa\,L}\,t\right)\=
\\[1ex]
&\=E_0
   -E_0\left(\frac{L}{\kappa}+\frac{\eta\,L}{2\sqrt{2}\,\pi\,\kappa}\right)\,
   \frac{2\sqrt{2}\,\pi\,\kappa}{\eta\,L}\,\exp\left(-\frac{2\sqrt{2}\,\pi\,\kappa}{\eta\,L}\,t\right)\:.
\end{split}
\end{equation}

\section{Discussion}
\sloppy
The solution of the differential equations for the vector potential gives rise to two very interesting situations. The first one is connected with a non-zero value of the generalized $E_z$ component, determining the emergence of a new, detectable contribution along the direction of the magnetic field.
The value of this new electric field is expressed, in dimensional units, as
\begin{equation}
E_{z}\=\frac{4\,\pi\,\Bc(T)\,\mathcal{D}}{\eta\,L}\:.
\end{equation}
For an YBCO sample of thickness $L=10\,\cm$ at a temperature $T=85\,\text{K}$, this would correspond to a value $E_{z}\simeq10\,\text{V}/\text{m}$.\par
\begin{figure}[!htp]
\captionsetup{skip=5pt,belowskip=15pt,font=small,labelfont=small,format=hang}
\centering
\includegraphics[width=0.95\textwidth,keepaspectratio]{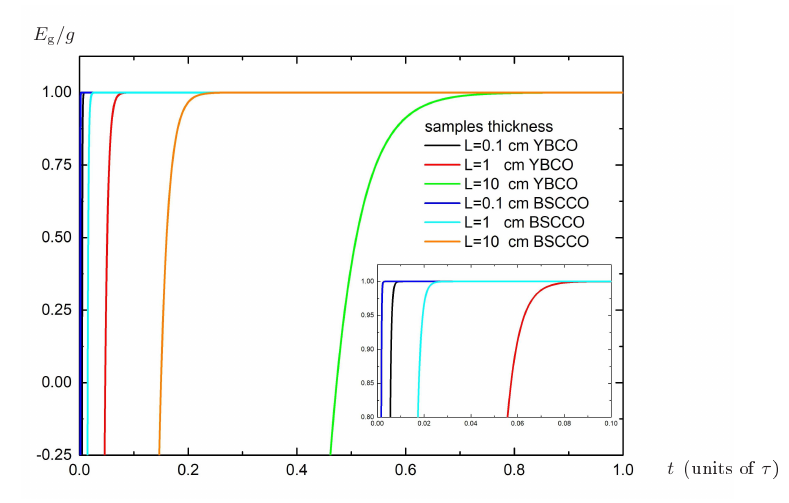}
\caption{Variation of gravitational field as a function of time inside superconductive samples of YBCO (black, red and green solid line, $T=85\,\Kelv$) and BSCCO (dark blue, blue and orange solid line, $T=102\,\Kelv$) for different thickness values ($L=0.1\,\text{cm}$, \,$L=1\,\text{cm}$, \,$L=10\,\text{cm}$). In the inset, the same results are plotted in a different scale (same axes labels), to better appreciate the variations at smaller times.}
\label{fig:graphs}
\end{figure}

The second remarkable effect is related to the expected variation of the gravitational field along the $x$ direction inside the superconductor. In fact, it is possible to see from eq. \eqref{eq:Excorr} and from Fig.\ \ref{fig:graphs} that the field is reduced with respect to its external value. Moreover, in analogy to what we found in \cite{Ummarino:2019cvw}, for very short time scales ($\sim 10^{-9}\,\text{s}$), the gravitational field seems to change sign: this could only happen in absence of suitable physical cutoffs preventing arbitrary growth of instabilities, giving rise to negative values \cite{Modanese:1995tx}.
It can also be noted that it is possible to find observable effects of gravitational field affection in different time scales; in principle, when dealing with samples of larger thickness and at temperatures very close to $\Tc$, larger observation times could be achieved.\par
In order to measure the effect it is necessary to determine suitable sample dimensions and chemical composition. In fact, large dimensions of the sample give rise to an increase for the time scales in which the effect occurs, combined with a decreasing intensity of the phenomenon. At the same time, large values of the $\eta$ parameter (related to the sample characteristics) determines similar effect, that is, a reduced intensity of the correction together with an increase of the observation times. In this regard, the $L$ and $\eta$ parameters act in the same way: disordered materials (small $\eta$ values, i.e.\ bad conductors in the normal state) with negligible thickness dimension (small $L$ values) give rise to larger effects for a shorter time%
\footnote{%
this would be the case, for example, of superconducting films with high disorder}.
On the contrary, large dimensions materials with good conductivity in the normal state (large $\eta$), would exhibit a weaker effect for a longer time scale. All this observations are in agreement with what we found in \cite{Ummarino:2019cvw}.\par
In Table \ref{tab:YBCOvsBSSCO} we report typical values for the physical parameters of two common high-$\Tc$ cuprates, YBCO and BSCCO \cite{Ummarino:2017bvz,weigand2010mixed,piriou2008effect,camargo2014first}. We also point out that the presence of disorder also determines an increase of the $\lambda$ penetration depth and, being the square of the latter proportional to the time scale ($\tau\propto\lambda^2$), the duration of the phenomenon increases with beneficial effects for direct measurements. Finally, if we put ourselves at temperatures very close to $\Tc$, we again find an increase in the $\lambda$ values together with larger time scales; in the latter case, however, the effects of thermal fluctuations should also be considered \cite{larkin2005theory}.\par
We expect that experimental issues would reside in the very short observation time. Since even detecting a reduced effect would be a remarkable result, it would be better to privilege long time scales setup, rather than configurations leading to larger effects of affection of the gravitational field. The latter would in fact imply strong difficulties in the measurements, since they would manifest themselves only for very short time scales. In light of these considerations, we suggest to consider big samples featuring large $\eta$ values.
%

\section{Concluding remarks}
A deeper interweaving between condensed matter and gravitational theories has already proven to be a powerful tool for inspecting many aspects of fundamental physics \cite{Anandan:1985pv,Zurek:1996sj,Jacobson:1998ms,Volovik:2000ua,Kiefer:2004hv,
Minter:2009fx,Quach:2015qwa,zaanen2015holographic,Bagchi:2021avw}, while also providing new insights into related unresolved issues%
\footnote{%
see for example `analogue gravity' techniques exploiting a bottom-up formulation for condensed matter systems featuring analogues of gravitational effects \cite{Garay:1999sk,Barcelo:2000tg,novello2002artificial,Barcelo:2005fc,
Carusotto:2008ep,Boada:2010sh,Gallerati:2018dgm,Capozziello:2018mqy,
Franz:2018cqi,Hu:2018psq,Gallerati:2021rtp,Kolobov:2021ynv}, or top-down holographic approaches, where a substrate description comes from the geometric formulation of a suitable gravitational model \cite{Andrianopoli:2019sip,Gallerati:2021htm}.
}.\par
In this paper we have exploited a multidisciplinary approach to describe a gravity/superfluid interplay, in a specific physical setup involving the presence of external electric and magnetic fields, which in turn determine the presence of a vortex lattice. This situation could induce enhanced effects in the proposed interplay, leading to a non-negligible local affection of the gravitational field within the condensate, together with a predicted electric field generated inside the sample, parallel to the superconductor plane. The experimental verification of the emergence of this new component would result in a great step forward in the study of the interaction of the gravitational field with quantum condensates, opening new and unsuspected horizons both in the theoretical and applicative fields.

\begin{table}[!htp]
\centering
\makegapedcells
\setcellgapes{5pt} 
\begin{tabular}
{@{} M{p}{0.15} M{p}{0.175} M{p}{0.25} @{}}
\toprule
\midrule
                  &  \text{YBCO}                 &  \text{BSCCO}  \\
\thinrule
\Tc               &  89\,\Kelv                   &  107\,\Kelv \\
T_\star           &  85\,\Kelv                   &  102\,\Kelv \\
\xi(T_\star)      &  8.49\cdot10^{-9}\,\mt       &  4.63\cdot10^{-9}\,\mt \\
\lambda(T_\star)  &  8.02\cdot10^{-7}\,\mt       &  1.11\cdot10^{-6}\,\mt \\
\sigma^{-1}       &  4.0\cdot10^{-7}\,\Omega\,\mt\,{}^\ms{(\ast)}
                                                 &
                                           3.6\cdot10^{-6}~\Omega\,\mt\,{}^\ms{(\ast\ast)}\\
\Bctwo(0)         &  61\,\mathrm{Tesla}          &  113\,\mathrm{Tesla}\\
\Bctwo(T_\star)   &  6\,\mathrm{Tesla}           &  11\,\mathrm{Tesla}\\
\Bc(T_\star)      &  0.25\,\mathrm{Tesla}        &  0.32~\mathrm{Tesla}\\
\lambda(0)        &  1.7\cdot10^{-7}\,\mt        &  2.4\cdot10^{-7}\,\mt\\
\xi(0)            &  1.8\cdot10^{-9}\,\mt        &  10^{-9}\,\mt\\
\kappa            &  94.4                        &  240.0 \\
\tau(T_\star)     &  2.01\cdot10^{-9}\,\s        &  1.23\cdot10^{-9}\,\s \\
\eta              &  1.0\cdot10^{-3}             &  3.5\cdot10^{-4} \\
\mathcal{D}       &  3.2\cdot10^{-4}\,\mt^{2}/\s &  10^{-3}\,\mt^{2}/\s\\
{}                &  \text{\fns${}^\ms{(\ast)}\;T\,=\,90\,\Kelv$}
                                                 &\text{\fns${}^\ms{(\ast\ast)}\;T\,=\,108\,\Kelv$}\\
\midrule
\bottomrule
\end{tabular}
\caption{%
YBCO vs.\ BSCCO \cite{Ummarino:2017bvz,weigand2010mixed,piriou2008effect,camargo2014first}
}
\label{tab:YBCOvsBSSCO}
\end{table}


\bigskip
\section*{\normalsize Acknowledgments}
\vspace{-5pt}
G.A. Ummarino acknowledges partial support from the MEPhI.
We also thank Fondazione CRT \,\includegraphics[height=\fontcharht\font`\B]{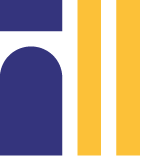}\:
that partially supported this work for A.\ Gallerati.

%


\pagebreak
\hypersetup{linkcolor=blue}
\phantomsection 
\addtocontents{toc}{\protect\addvspace{3.5pt}}
\addcontentsline{toc}{section}{References} 
\bibliographystyle{mybibstyle}
\bibliography{bibliografia} 

\providecommand{\href}[2]{#2}\begingroup\begin{thebibliography}{10}

\bibitem{DeWitt:1966yi}
Bryce~S. DeWitt, \textit{``{Superconductors and gravitational drag}''}, Phys.\
  Rev.\ Lett.\ \textbf{16} (1966) 1092--1093.

\bibitem{papini1967detection}
G.~Papini, \textit{``{Detection of inertial effects with superconducting
  interferometers}''}, Physics Letters A \textbf{24} (1967), n.~1, 32--33.

\bibitem{Papini:1970cw}
G.~Papini, \textit{``{Superconducting and normal metals as detectors of
  gravitational waves}''}, Lett. Nuovo Cim. \textbf{4S1} (1970) 1027--1032.

\bibitem{hirakawa1975superconductors}
H.~Hirakawa, \textit{``Superconductors in gravitational field''}, Physics
  Letters A \textbf{53} (1975), n.~5, 395--396.

\bibitem{anandan1984relthe}
J.~Anandan, \textit{``Relativistic thermoelectromagnetic gravitational effects
  in normal conductors and superconductors''}, Physics Letters A \textbf{105}
  (1984), n.~6, 280--284.

\bibitem{ross1983london}
D.K. Ross, \textit{``{The London equations for superconductors in a
  gravitational field}''}, Journal of Physics A: Mathematical and General
  \textbf{16} (1983), n.~6, 1331.

\bibitem{dinariev1987relativistic}
O.~Yu Dinariev and A.B. Mosolov, \textit{``A relativistic effect in the theory
  of superconductivity''}, Dokl.\ Akad.\ Nauk SSSR \textbf{295} (1987) 98.

\bibitem{peng1990new}
H.~Peng, \textit{``A new approach to studying local gravitomagnetic effects on
  a superconductor''}, General Relativity and Gravitation \textbf{22} (1990),
  n.~6, 609--617.

\bibitem{peng1991electrodynamics}
H.~Peng, D.G. Torr, E.K. Hu and B.~Peng, \textit{``Electrodynamics of moving
  superconductors and superconductors under the influence of external
  forces''}, Phys. Rev. B \textbf{43} (1991), n.~4, 2700.

\bibitem{peng1991interaction}
H.~Peng, G.~Lind and Y.S. Chin, \textit{``Interaction between gravity and
  moving superconductors''}, General relativity and gravitation \textbf{23}
  (1991), n.~11, 1231--1250.

\bibitem{li1991effects}
Ning Li and D.G. Torr, \textit{``Effects of a gravitomagnetic field on pure
  superconductors''}, Physical Review D \textbf{43} (1991), n.~2, 457.

\bibitem{podkletnov1992possibility}
E.~Podkletnov and R.~Nieminen, \textit{``A possibility of gravitational force
  shielding by bulk \textrm{YBa${}_2$Cu${}_3$O${}_{7-\mathrm{X}}$}
  superconductor''}, Physica C: Superconductivity \textbf{203} (1992), n.~3-4,
  441--444.

\bibitem{podkletnov1997weak}
E.~Podkletnov, \textit{``{Weak gravitation shielding properties of composite
  bulk \textrm{YBa${}_2$Cu${}_3$O${}_{7-\mathrm{X}}$} superconductor below 70 K
  under EM field}''}, cond-mat/9701074 (1997).

\bibitem{hathaway2003gravity}
G.~Hathaway, B.~Cleveland and Y.~Bao, \textit{``Gravity modification experiment
  using a rotating superconducting disk and radio frequency fields''}, Physica
  C: Superconductivity \textbf{385} (2003), n.~4, 488.

\bibitem{li1997static}
Ning Li, David Noever, Tony Robertson, Ron Koczor and Whitt Brantley,
  \textit{``Static test for a gravitational force coupled to type \textrm{II
  YBCO} superconductors''}, Physica C: Superconductivity \textbf{281} (1997),
  n.~2, 260--267.

\bibitem{Modanese:2001wv}
Giovanni Modanese, \textit{``{Local contribution of a quantum condensate to the
  vacuum energy density}''}, Mod. Phys. Lett. A \textbf{18} (2003) 683--690,
  [\href{http://arxiv.org/abs/gr-qc/0107073}{\texttt{gr-qc/0107073}}].

\bibitem{Modanese:1996sq}
Giovanni Modanese, \textit{``{Possible quantum gravity effects in a charged
  Bose condensate under variable em field}''}, Phys. Essays \textbf{14} (2001)
  93--105, [\href{http://arxiv.org/abs/gr-qc/9612022}{\texttt{gr-qc/9612022}}].

\bibitem{Wu:2003aq}
Ning Wu, \textit{``{Gravitational shielding effects in gauge theory of
  gravity}''}, Commun. Theor. Phys. \textbf{41} (2004) 567--572,
  [\href{http://arxiv.org/abs/hep-th/0307225}{\texttt{hep-th/0307225}}].

\bibitem{Modanese:1995tx}
Giovanni Modanese, \textit{``{Theoretical analysis of a reported weak
  gravitational shielding effect}''}, Europhys.\ Lett. \textbf{35} (1996)
  413--418,
  [\href{http://arxiv.org/abs/hep-th/9505094}{\texttt{hep-th/9505094}}].

\bibitem{Modanese:1996zm}
Giovanni Modanese, \textit{``{Role of a `local' cosmological constant in
  Euclidean quantum gravity}''}, Phys. Rev. D \textbf{54} (1996) 5002--5009,
  [\href{http://arxiv.org/abs/hep-th/9601160}{\texttt{hep-th/9601160}}].

\bibitem{schiff1966gravitation}
L.I. Schiff and M.V. Barnhill, \textit{``Gravitation-induced electric field
  near a metal''}, Physical Review \textbf{151} (1966), n.~4, 1067.

\bibitem{Dessler:1968zz}
A.J. Dessler, F.C. Michel, H.E. Rorschach and G.T. Trammell,
  \textit{``{Gravitationally Induced Electric Fields in Conductors}''}, Phys.
  Rev. \textbf{168} (1968) 737--743.

\bibitem{torr1993gravitoelectric}
Douglas~G. Torr and Ning Li, \textit{``Gravitoelectric-electric coupling via
  superconductivity''}, Foundations of Physics Letters \textbf{6} (1993), n.~4,
  371--383.

\bibitem{mashhoon2000gravitoelectromagnetism}
Bahram Mashhoon, \textit{``{Gravitoelectromagnetism}''}, in {\em {Reference
  Frames and Gravitomagnetism}}, pp.~121--132, World Scientific Publishing,
  Singapore (2000).

\bibitem{mashhoon2007review}
Bahram Mashhoon, \textit{``Gravitoelectromagnetism: A brief review''}, in {\em
  The Measurement of Gravitomagnetism: A Challenging Enterprise}, ch.~3,
  pp.~29--40, Nova Science Publisher Inc., New York, USA (2007).

\bibitem{witteborn1967experimental}
F.C. Witteborn and W.M. Fairbank, \textit{``Experimental comparison of the
  gravitational force on freely falling electrons and metallic electrons''},
  Physical Review Letters \textbf{19} (1967), n.~18, 1049.

\bibitem{witteborn1968experiments}
F.C. Witteborn and W.M. Fairbank, \textit{``Experiments to determine the force
  of gravity on single electrons and positrons''}, Nature \textbf{220} (1968),
  n.~5166, 436--440.

\bibitem{herring1968gravitationally}
C.~Herring, \textit{``Gravitationally induced electric field near a conductor,
  and its relation to the surface-stress concept''}, Phys. Rev. \textbf{171}
  (1968), n.~5, 1361.

\bibitem{agop2000local}
M.~Agop, C.Gh. Buzea and P.~Nica, \textit{``Local gravitoelectromagnetic
  effects on a superconductor''}, Physica C: Superconductivity \textbf{339}
  (2000), n.~2, 120--128.

\bibitem{agop2000some}
M.~Agop, P.D. Ioannou and F.~Diaconu, \textit{``Some implications of
  gravitational superconductivity''}, Progress of Theoretical Physics
  \textbf{104} (2000), n.~4, 733--742.

\bibitem{Ruggiero:2002hz}
Matteo~Luca Ruggiero and Angelo Tartaglia, \textit{``{Gravitomagnetic
  effects}''}, Nuovo Cim. B \textbf{117} (2002) 743--768,
  [\href{http://arxiv.org/abs/gr-qc/0207065}{\texttt{gr-qc/0207065}}].

\bibitem{Tartaglia:2003wx}
Angelo Tartaglia and Matteo~Luca Ruggiero, \textit{``{Gravitoelectromagnetism
  versus electromagnetism}''}, Eur. J. Phys. \textbf{25} (2004) 203--210,
  [\href{http://arxiv.org/abs/gr-qc/0311024}{\texttt{gr-qc/0311024}}].

\bibitem{Ummarino:2017bvz}
Giovanni~Alberto Ummarino and Antonio Gallerati, \textit{``{Superconductor in a
  weak static gravitational field}''}, Eur. Phys. J. \textbf{C77} (2017), n.~8,
  549, [\href{http://arxiv.org/abs/1710.01267}{\texttt{arXiv:1710.01267}}].

\bibitem{Behera:2017voq}
Harihar Behera, \textit{``{Comments on gravitoelectromagnetism of Ummarino and
  Gallerati in “Superconductor in a weak static gravitational field” vs
  other versions}''}, Eur. Phys. J. \textbf{C77} (2017), n.~12, 822,
  [\href{http://arxiv.org/abs/1709.04352}{\texttt{arXiv:1709.04352}}].

\bibitem{Ummarino:2019cvw}
Giovanni~Alberto Ummarino and Antonio Gallerati, \textit{``{Exploiting weak
  field gravity-Maxwell symmetry in superconductive fluctuations regime}''},
  Symmetry \textbf{11} (2019), n.~11, 1341,
  [\href{http://arxiv.org/abs/1910.13897}{\texttt{arXiv:1910.13897}}].

\bibitem{Ummarino:2020loo}
Giovanni~Alberto Ummarino and Antonio Gallerati, \textit{``{Josephson AC effect
  induced by weak gravitational field}''}, Class. Quant. Grav. \textbf{37}
  (2020), n.~21, 217001,
  [\href{http://arxiv.org/abs/2009.04967}{\texttt{arXiv:2009.04967}}].

\bibitem{Ummarino:2021vwc}
G.~A. Ummarino and A.~Gallerati, \textit{``{Possible alterations of local
  gravitational field inside a superconductor}''}, Entropy \textbf{23} (2021),
  n.~2, 193,
  [\href{http://arxiv.org/abs/2102.01489}{\texttt{arXiv:2102.01489}}].

\bibitem{Gallerati:2021ops}
Antonio Gallerati, \textit{``{Local affection of weak gravitational field from
  supercondensates}''}, Phys. Scripta \textbf{96} (2021), n.~6, 064001.

\bibitem{Vieira:2016csi}
R.~S. Vieira and H.~B. Brentan, \textit{``{Covariant theory of gravitation in
  the framework of special relativity}''}, Eur. Phys. J. Plus \textbf{133}
  (2018) 165,
  [\href{http://arxiv.org/abs/1608.00815}{\texttt{arXiv:1608.00815}}].

\bibitem{Sbitnev:2019iyz}
Valeriy~I. Sbitnev, \textit{``{Quaternion algebra on 4D superfluid quantum
  space-time. Gravitomagnetism}''}, Found. Phys. \textbf{49} (2019), n.~2,
  107--143,
  [\href{http://arxiv.org/abs/1901.09098}{\texttt{arXiv:1901.09098}}].

\bibitem{Giardino:2018ffd}
Sergio Giardino, \textit{``{A novel covariant approach to
  gravito-electromagnetism}''}, Braz. J. Phys. \textbf{50} (2020), n.~3,
  372--378,
  [\href{http://arxiv.org/abs/1812.07371}{\texttt{arXiv:1812.07371}}].

\bibitem{Gallerati:2020tyq}
Antonio Gallerati, \textit{``{Interaction between superconductors and weak
  gravitational field}''}, J. Phys. Conf. Ser. \textbf{1690} (2020), n.~1,
  012141, [\href{http://arxiv.org/abs/2101.00418}{\texttt{arXiv:2101.00418}}].

\bibitem{tang1995time}
Q.~Tang and S~Wang, \textit{``{Time dependent Ginzburg-Landau equations of
  superconductivity}''}, Physica D: Nonlinear Phenomena \textbf{88} (1995),
  n.~3-4, 139--166.

\bibitem{lin1997ginzburg}
Fang-Hua Lin and Qiang Du, \textit{``{Ginzburg-Landau vortices: dynamics,
  pinning, and hysteresis}''}, SIAM Journal on Mathematical Analysis
  \textbf{28} (1997), n.~6, 1265--1293.

\bibitem{ullah1991effect}
S.~Ullah and A.T. Dorsey, \textit{``Effect of fluctuations on the transport
  properties of type-ii superconductors in a magnetic field''}, Physical Review
  B \textbf{44} (1991), n.~1, 262.

\bibitem{ghinovker1999explosive}
M.~Ghinovker, I.~Shapiro and B.~Ya Shapiro, \textit{``Explosive nucleation of
  superconductivity in a magnetic field''}, Physical Review B \textbf{59}
  (1999), n.~14, 9514.

\bibitem{kopnin1999time}
N.B. Kopnin and E.V. Thuneberg, \textit{``{Time-dependent Ginzburg-Landau
  analysis of inhomogeneous normal-superfluid transitions}''}, Phys. Rev. Lett.
  \textbf{83} (1999), n.~1, 116.

\bibitem{fleckinger1998dynamics}
Jacqueline Fleckinger-Pell{\'e}, Hans~G Kaper and Peter Tak{\'a}{\v{c}},
  \textit{``{Dynamics of the Ginzburg-Landau equations of
  superconductivity}''}, Nonlinear Analysis: Theory, Methods \& Applications
  \textbf{32} (1998), n.~5, 647--665.

\bibitem{du1996high}
Qiang Du and Paul Gray, \textit{``{High-kappa limits of the time-dependent
  Ginzburg-Landau model}''}, SIAM Journal on Applied Mathematics \textbf{56}
  (1996), n.~4, 1060--1093.

\bibitem{oripov2020time}
Bakhrom Oripov and Steven~M Anlage, \textit{``{Time-dependent Ginzburg-Landau
  treatment of rf magnetic vortices in superconductors: Vortex semiloops in a
  spatially nonuniform magnetic field}''}, Phys. Rev. E \textbf{101} (2020),
  n.~3, 033306.

\bibitem{kopnin1993flux}
N.B. Kopnin, B.I. Ivlev and V.A. Kalatsky, \textit{``{The flux-flow Hall effect
  in type II superconductors. An explanation of the sign reversal}''}, J. Low
  Temp. Phys. \textbf{90} (1993), n.~1, 1--13.

\bibitem{kopnin2001theory}
Nikolai Kopnin, \textit{``Theory of nonequilibrium superconductivity''}; Oxford
  University Press, Oxford, UK (2001).

\bibitem{ketterson1999superconductivity}
J.B. Ketterson and S.N. Song, \textit{``Superconductivity''}; Cambridge
  University Press, Cambridge, UK (1999).

\bibitem{hoffmann2012ginzburg}
K.H. Hoffmann and Qi~Tang, \textit{``{Ginzburg-Landau phase transition theory
  and superconductivity}''}; Springer Basel AG, Basel, Switzerland (2012).

\bibitem{sanders2007averaging}
Jan~A. Sanders, Ferdinand Verhulst and James Murdock, \textit{``Averaging
  methods in nonlinear dynamical systems''}; Springer Verlag, New York, USA
  (2007).

\bibitem{weigand2010mixed}
M.~Weigand, M.~Eisterer, E.~Giannini and H.W. Weber, \textit{``{Mixed state
  properties of Bi${}_2$Sr${}_2$Ca${}_2$Cu${}_3$O${}_{10+\delta}$ single
  crystals before and after neutron irradiation}''}, Phys. Rev. B \textbf{81}
  (2010), n.~1, 014516.

\bibitem{piriou2008effect}
A.~Piriou, Y.~Fasano, E.~Giannini and {\O}.~Fischer, \textit{``{Effect of
  oxygen-doping on Bi${}_2$Sr${}_2$Ca${}_2$Cu${}_3$O${}_{10+\delta}$ vortex
  matter: crossover from electromagnetic to Josephson interlayer coupling}''},
  Phys. Rev. B \textbf{77} (2008), n.~18, 184508.

\bibitem{camargo2014first}
J.A. Camargo-Mart{\'\i}nez, D.~Espitia and R.~Baquero,
  \textit{``{First-principles study of electronic structure of
  Bi${}_2$Sr${}_2$Ca${}_2$Cu${}_3$O${}_{10}$}''}, Revista Mexicana de
  F{\'\i}sica \textbf{60} (2014), n.~1, 39--45.

\bibitem{larkin2005theory}
Anatoli Larkin and Andrei Varlamov, \textit{``{Theory of fluctuations in
  superconductors}''}; Oxford University Press, Oxford, UK (2005).

\bibitem{Anandan:1985pv}
J.~Anandan, \textit{``{Detection of gravitational radiation using
  superconductors}''}, Phys. Lett. A \textbf{110} (1985) 446--450.

\bibitem{Zurek:1996sj}
W.~H. Zurek, \textit{``{Cosmological experiments in condensed matter
  systems}''}, Phys. Rept. \textbf{276} (1996) 177--221,
  [\href{http://arxiv.org/abs/cond-mat/9607135}{\texttt{cond-mat/9607135}}].

\bibitem{Jacobson:1998ms}
T.~A. Jacobson and G.~E. Volovik, \textit{``{Event horizons and ergoregions in
  He-3}''}, Phys. Rev. D \textbf{58} (1998) 064021,
  [\href{http://arxiv.org/abs/cond-mat/9801308}{\texttt{cond-mat/9801308}}].

\bibitem{Volovik:2000ua}
G.~E. Volovik, \textit{``{Superfluid analogies of cosmological phenomena}''},
  Phys. Rept. \textbf{351} (2001) 195--348,
  [\href{http://arxiv.org/abs/gr-qc/0005091}{\texttt{gr-qc/0005091}}].

\bibitem{Kiefer:2004hv}
Claus Kiefer and Carsten Weber, \textit{``{On the interaction of mesoscopic
  quantum systems with gravity}''}, Annalen Phys. \textbf{14} (2005) 253--278,
  [\href{http://arxiv.org/abs/gr-qc/0408010}{\texttt{gr-qc/0408010}}].

\bibitem{Minter:2009fx}
Stephen~J. Minter, Kirk Wegter-McNelly and Raymond~Y. Chiao, \textit{``{Do
  Mirrors for Gravitational Waves Exist?}''}, Physica E \textbf{42} (2010) 234,
  [\href{http://arxiv.org/abs/0903.0661}{\texttt{arXiv:0903.0661}}].

\bibitem{Quach:2015qwa}
James~Q. Quach, \textit{``{Gravitational Casimir effect}''}, Phys. Rev. Lett.
  \textbf{114} (2015), n.~8, 081104,
  [\href{http://arxiv.org/abs/1502.07429}{\texttt{arXiv:1502.07429}}].
  [Erratum: Phys. Rev. Lett. \textbf{118} (2017) 139901].

\bibitem{zaanen2015holographic}
{Zaanen, Jan and Liu, Yan and Sun, Ya-Wen and Schalm, Koenraad},
  \textit{``{Holographic duality in condensed matter physics}''}; {Cambridge
  University Press}, {Cambridge, UK} ({2015}).

\bibitem{Bagchi:2021avw}
Bijan Bagchi and Rahul Ghosh, \textit{``{Dirac Hamiltonian in a supersymmetric
  framework}''}, Journal of Mathematical Physics \textbf{62} (2021), n.~7,
  072101, [\href{http://arxiv.org/abs/2101.03922}{\texttt{arXiv:2101.03922}}].

\bibitem{Garay:1999sk}
L.~J. Garay, J.~R. Anglin, J.~I. Cirac and P.~Zoller, \textit{``{Black holes in
  Bose-Einstein condensates}''}, Phys. Rev. Lett. \textbf{85} (2000)
  4643--4647,
  [\href{http://arxiv.org/abs/gr-qc/0002015}{\texttt{gr-qc/0002015}}].

\bibitem{Barcelo:2000tg}
Carlos Barcelo, Stefano Liberati and Matt Visser, \textit{``{Analog gravity
  from Bose-Einstein condensates}''}, Class. Quant. Grav. \textbf{18} (2001)
  1137, [\href{http://arxiv.org/abs/gr-qc/0011026}{\texttt{gr-qc/0011026}}].

\bibitem{novello2002artificial}
{Novello, Mario and Visser, Matt and Volovik, Grigory E.},
  \textit{``{Artificial black holes}''}; {World Scientific}, {Singapore}
  ({2002}).

\bibitem{Barcelo:2005fc}
Carlos Barcelo, Stefano Liberati and Matt Visser, \textit{``{Analogue
  gravity}''}, Living Rev. Rel. \textbf{8} (2005) 12.

\bibitem{Carusotto:2008ep}
Iacopo Carusotto, Serena Fagnocchi, Alessio Recati, Roberto Balbinot and
  Alessandro Fabbri, \textit{``{Numerical observation of Hawking radiation from
  acoustic black holes in atomic Bose-Einstein condensates}''}, New J. Phys.
  \textbf{10} (2008) 103001,
  [\href{http://arxiv.org/abs/0803.0507}{\texttt{arXiv:0803.0507}}].

\bibitem{Boada:2010sh}
O.~Boada, A.~Celi, J.~I. Latorre and M.~Lewenstein, \textit{``{Dirac Equation
  For Cold Atoms In Artificial Curved Spacetimes}''}, New J. Phys. \textbf{13}
  (2011) 035002,
  [\href{http://arxiv.org/abs/1010.1716}{\texttt{arXiv:1010.1716}}].

\bibitem{Gallerati:2018dgm}
Antonio Gallerati, \textit{``{Graphene properties from curved space Dirac
  equation}''}, Eur. Phys. J. Plus \textbf{134} (2019), n.~5, 202,
  [\href{http://arxiv.org/abs/1808.01187}{\texttt{arXiv:1808.01187}}].

\bibitem{Capozziello:2018mqy}
Salvatore Capozziello, Richard Pincak and Emmanuel~N. Saridakis,
  \textit{``{Constructing superconductors by graphene Chern-Simons
  wormholes}''}, Annals Phys. \textbf{390} (2018) 303--333.

\bibitem{Franz:2018cqi}
M.~Franz and M.~Rozali, \textit{``{Mimicking black hole event horizons in
  atomic and solid-state systems}''}, Nature Rev. Mater. \textbf{3} (2018)
  491--501,
  [\href{http://arxiv.org/abs/1808.00541}{\texttt{arXiv:1808.00541}}].

\bibitem{Hu:2018psq}
Jiazhong Hu, Lei Feng, Zhendong Zhang and Cheng Chin, \textit{``{Quantum
  simulation of Unruh radiation}''}, Nature Phys. \textbf{15} (2019), n.~8,
  785--789,
  [\href{http://arxiv.org/abs/1807.07504}{\texttt{arXiv:1807.07504}}].

\bibitem{Gallerati:2021rtp}
Antonio Gallerati, \textit{``{Negative-curvature spacetime solutions for
  graphene}''}, J. Phys. Condens. Matter \textbf{33} (2021), n.~13, 135501,
  [\href{http://arxiv.org/abs/2101.03010}{\texttt{arXiv:2101.03010}}].

\bibitem{Kolobov:2021ynv}
Victor~I. Kolobov, Katrine Golubkov, Juan~Ram\'on Mu\~noz~de Nova and Jeff
  Steinhauer, \textit{``{Observation of stationary spontaneous Hawking
  radiation and the time evolution of an analogue black hole}''}, Nature Phys.
  \textbf{17} (2021), n.~3, 362--367.

\bibitem{Andrianopoli:2019sip}
L.~Andrianopoli, B.~L. Cerchiai, R.~D'Auria, A.~Gallerati, R.~Noris,
  M.~Trigiante and J.~Zanelli, \textit{``{$\mathcal{N}$-extended $D = 4$
  supergravity, unconventional SUSY and graphene}''}, JHEP \textbf{01} (2020)
  084, [\href{http://arxiv.org/abs/1910.03508}{\texttt{arXiv:1910.03508}}].

\bibitem{Gallerati:2021htm}
A.~Gallerati, \textit{``{Supersymmetric theories and graphene}''}, PoS
  \textbf{390} (2021) 662,
  [\href{http://arxiv.org/abs/2104.07420}{\texttt{arXiv:2104.07420}}].

\end{thebibliography}\endgroup


\end{document}